\documentclass[prd,aps,preprint,epsfig,floats]{revtex4}

\usepackage{graphicx}

\def\dsodt{ds_{23}\over dt}
\def\dstdt{ds_{13}\over dt}
\def\dsthdt{ds_{12}\over dt}

\def\be{\begin{equation}}
\def\ee{\end{equation}}
\def\ba{\begin{eqnarray}}
\def\ea{\end{eqnarray}}
\def\br{\begin{array}}
\def\er{\end{array}}

\input epsf.tex
\def\DESepsf(#1 width #2){\epsfxsize=#2 \epsfbox{#1}}

\begin{document}

\renewcommand{\thefootnote}{\alph{footnote}}


\title{{\bf  CAN MEASUREMENT OF $\theta_{13}$ TELL US ABOUT 
QUARK-LEPTON UNIFICATION ?}}
\author{\bf  R.N. MOHAPATRA}

\address{ Department of Physics, University of Maryland, College Park,
MD-20742, USA}

\begin{abstract} 

We argue that a high precision measurement of the 
neutrino mixing
parameter $\theta_{13}$, within a three neutrino seesaw framework can
throw important light on the question of whether the quarks and leptons
unify into a single matter at high scales. Based on a number of
examples, we conclude that a value of $\theta_{13} \leq {\frac{\Delta
m^2_{\odot}}{\Delta m^2_A}}\simeq 0.04$ would require at the minimum a 
pure leptonic
interchange symmetry between $\mu$ and $\tau$ generations for its natural 
theoretical understanding and will
disfavor a quark-lepton unification type theory such as those based on
$SU(4)_c$ or $SO(10)$ whereas a bigger value would leave open the
possibility of quark lepton unification.
 \end{abstract}

\maketitle

\vskip1.0in

\section{Introduction}
There are several reasons to think that quarks and lepton may be two
different manifestations of the same form of matter. The first hint arises
from the observed similarities between weak interaction properties
of quarks and leptons. This is generally known as quark-lepton symmetry
and even though it manifests itself only in the left-handed
helicity sector of quarks and leptons, it is often considered as a hint
of further unification among these two very different kinds of matter.
The second comes from the attractive hypothesis of
grand unification of matter and forces which argues that at very 
short distances, all forces and all matter unify. Even though there is no
direct experimental evidence for grand unification, the apparent  
couplings unification in a simple supersymmetric model keeps this as a
popular idea in the modern particle theory. 

There are of course many observations that provide strong distinctions 
between quarks and leptons:  for instance,

\begin{itemize}

\item  Quarks have strong interactions whereas leptons do
not.
\item  The standard model has both up and down right
handed quarks experiencing
the gauge forces whereas in the lepton sector only the right handed 
charged lepton (which is the analog of the down type quark) experiences
the gauge force and the right handed neutrino  which is the analog of the 
up quark does not.

\item The mixing pattern among quarks of different
generations is very different from that among the leptons. For the quarks, 
we have the ``nearest neighbour rule'' i.e. $\theta_{12}\simeq 13^o$,  $
\theta_{23} \simeq 2^o$ and $ \theta_{13}\simeq 0.3^o$ whereas for the 
leptons one has
$\theta_{23} \simeq 45^o$; $ \theta_{12}\simeq 32^o$ whereas 
$\theta_{13}\leq 12^o$. 

\end{itemize}
 
One might therefore conclude that, while it is tempting
to speculate about quark lepton unification, the above experimentally 
observed facts
rule out such a possibility. Such conclusion however may be quite 
premature, since it could very easily be that quark-lepton unification is
not
manifest at low energies but could be present at very high energies.
In fact there exist very compelling theoretical frameworks that unify
quarks and leptons and yet are fully compatible
with the above observations.  
It is then important to look for low energy signals for such unification.

In this talk, I will argue that there may exist such a signal in the
neutrino sector. In order to advance my argument, I need to specify the
theretical framework within which this conclusion is valid. I will make
the minimal set of assumptions, which are quite plausible and are popular
in the discussion of neutrino mixings.

\bigskip

 (i) There are three
generations of neutrinos and they are Majorana fermions.

\bigskip

 (ii) The smallness
of neutrino masses is explained by the seesaw mechanism\cite{seesaw}.

\bigskip

 (iii) The high scale
quark-lepton unification theory is based on the Pati-Salam $SU(4)_c$ 
group or SO(10)\cite{ps}.

\bigskip

(iv) The charged leptons do not contribute significantly to neutrino 
mixings.

\bigskip

The assumption (ii) would in fact be required in some form if we want to
understand why the quark and lepton mixing angles are different in a Q-L
unified theory; since the different neutrino mixings could then be 
attributed to an apppropriate flavor structure among the right handed
neutrinos\cite{smirnov}. Within this set of assumptions, three examples
are presented where high scale quark-lepton unification implies
$\theta_{13}\sim \sqrt{\frac{\Delta
m^2_{\odot}}{\Delta m^2_A}}\simeq 0.1$ or so. I then argue that if
$\theta_{13}\leq \frac{\Delta
m^2_{\odot}}{\Delta m^2_A}\simeq 0.04$, then a fine tuning among the
different parameters of the theory is required and a natural way to
guarantee such small values is to have a purely leptonic symmetry
i.e. $\mu-\tau$ interchange, a symmetry not apparently shared by quarks. 
In such a
situation, it is very unlikely that there will be quark lepton unification
at high scale.

This talk is organized as follows: in sec. 2, seesaw mechanism is
discussed and general arguments are given in favor of the basic thesis of 
this
paper. In sec. 3, three models that embed quark-lepton unifying group
$SU(4)_c$ are shown to predict ``large'' $\theta_{13}$ as argued in
the introduction. In sec. 4, it shown how $\mu-\tau$ interchange symmetry
can lead to ``small'' values of $\theta_{13}$.  

\section{Seesaw enables neutrino mixings to be large without conflicting
with quark-lepton unification}
In this section, we discuss how seesaw mechanism avoids an obvious
conflict between quark lepton unification and the vastly different mixing
patterns between quarks and leptons. To see why such a conflict would even
be contemplated, let us work with the gauge group
$SU(2)_L\times SU(2)_R\times SU(4)_c$ subgroup\cite{ps} under which the
fermions transform as
follows:
\begin{eqnarray}
{\bf \Psi}~=~\left(\begin{array}{cccc}u_1 & u_2 & u_3 & \nu\\ d_1 & d_2
& d_3& e\end{array}\right)
\end{eqnarray}
This group clearly unifies quarks with leptons. Therefore in the
``symmetry limit'' one would have  $M_u~=~M_{\nu^D}$ and
$M_\ell~=~M_d$. Therefore one might suspect that quark and neutrino
mass matrices could be similar. On the hand,
it is well known that in a basis where the up quark mass matrix is
diagonal, the down quark mass matrix has the generic form:
\begin{eqnarray}
M_{d}~\approx ~m_{b}\left(\begin{array}{ccc}\lambda^4 &
\lambda^3
&\lambda^3\\ \lambda^3 & \lambda^2& \lambda^2 \\ \lambda^3 & \lambda^2 &
1\end{array}\right)
\end{eqnarray}
where $\lambda \sim 0.22$ (the Cabibbo angle) and the matrix elements of 
the above matrix are supposed to represent 
only the approximate orders of magnitude. On the other hand, for leptons 
in the basis where charged leptons are mass eigenstates, the neutrino
Majorana mass matrix is given by:
\begin{eqnarray}
{ M}_\nu~=~\frac{\sqrt{\Delta
m^2_A}}{2}\left(\begin{array}{ccc}c\epsilon
&b\epsilon &d\epsilon\\ b\epsilon & 1+a\epsilon & -1 \\
d\epsilon & -1 & 1+\epsilon\end{array}\right)
\end{eqnarray}
where we have assumed the neutrino mass hierarchy to be normal; $\epsilon
\simeq \sqrt{\frac{\Delta m^2_\odot}{\Delta
m^2_A}}\approx \lambda$ and parameters $a,b,c,d$ are of order one. 

Note however that if small neutrino masses
arise from the seesaw mechanism, then their mass matrices arise from
either of the following two formulae  depending on whether the theory has
asymptotic parity symmetry or
not and they are:
\begin{eqnarray}
{ M}_\nu~=~-M^T_DM^{-1}_RM_D
\end{eqnarray}
which is the type I seesaw formula where $M_R$ is the right handed
neutrino mass matrix. In theories with parity symmetry, one gets instead
the type II seesaw formula\cite{seesaw2}:
\begin{eqnarray}
{ M}_\nu~=~fv_L-h^T_\nu f^{-1}_Rh_\nu
\left(\frac{v^2_{wk}}{v_R}\right)
\end{eqnarray}
Therefore even though in the $SU(4)_c$ limit, the quark and lepton mass 
matrices are identical, one could arrange the structure of $M_R$
such that, the large neutrino mixings arise purely from that structure.
For instance, if we take $[M_R]^{-1}_{ij}~=(\mu_{ij})^{-1}$, then the
condition for near maximal atmospheric mixing is $
\frac{\mu_{22}}{m^2_c}\sim \frac{\mu_{33}}{m^2_t} \sim
\frac{\mu_{23}}{m_cm_t}$ and that for solar is
$\frac{\mu_{13}}{m_u}\sim \frac{\mu_{33}}{m_t}\cdot \sqrt\frac{\Delta
m^2_\odot}{\Delta m^2_A}$. Clearly, the above relations imply severe fine
tuning among the right handed neutrino masses. In any case,
 for this toy example we can calculate the neutrino mixings as
follows. We now get $U_\nu$ of the form:
\begin{eqnarray}
U_\nu~=~\pmatrix{c & s & \delta\cr \frac{s}{\sqrt{2}} &
-\frac{c}{\sqrt{2}} & \frac{1}{\sqrt{2}}\cr \frac{s}{\sqrt{2}} &
-\frac{c}{\sqrt{2}} & -\frac{1}{\sqrt{2}}}
\end{eqnarray}
with a small value for the parameter $\delta$. 
Since the observed neutrino mixing matrix  $U_{PMNS}~=~U^{\dagger}_\ell
U_\nu$, we need $U_\ell$, which can be obtained in the exact quark-lepton
symmetric limit to be $U_\ell~=~ U^{\dagger}_{CKM}$. From this, we find
that $\theta_{13}\simeq \sum_\alpha
[U^{\dagger}_{CKM}]_{1\alpha}[U_\nu]_{\alpha 3}\sim \delta +
\frac{\lambda}{\sqrt{2}}+\frac{\lambda^3 a}{\sqrt{2}}$ (where $a$ is of
order one). 
This leads to $\theta_{13}\simeq 0.15$ which by our
definition is ``large'' without any fine tuning of parameters.
While this is a toy model, we see below that all the examples with
quark-lepton symmetry that we have explored, something similar happens 
 leading to the generic prediction that $\theta_{13}$ is ``large''.

\section{Simple breakings of quark-lepton unification and
 ``large'' $\theta_{13}$}
In the above example, the quark-lepton symmetry is assumed to be
exact for the Yukawa couplings that lead to charged fermion masses. In
this section, we depart from this simple example and consider models with
simple breaking of $SU(4)_c$ in the Yukawa couplings that lead to charged
fermion masses and argue that in this case we also get $\theta_{13}$ to be
large.  
\subsection{An $SU(2)_L\times SU(2)_R\times  SU(4)_c$ example}
Let us consider a model with three sets of Higgs fields: $\phi(2,2,0)$,
$\Sigma (2,2,15)$ and $\Delta(3,1,10)\oplus \Delta^c(1,3,\bar{10})$.
The Yukawa superpotential of this model is:
\begin{eqnarray}
W~=~h\Psi\phi\Psi^c~+~f\Psi\Sigma\Psi^c~+~f_\nu (\Psi\Psi\Delta +
\Psi^c\Psi^c\Delta^c)
\end{eqnarray}
After spontaneous breakdown of the gauge symmetry, we get for the fermion
mass matrices
\begin{eqnarray}
M_u~=~ h \kappa_u + f v_u \\\nonumber
M_d~=~ h \kappa_d + f v_d \\  \nonumber
M_\ell~=~ h \kappa_d -3 f v_d \\  \nonumber
M_{\nu_D}~=~ h \kappa_u -3 f v_u \\\nonumber
\label{eq1}\end{eqnarray}
where $\kappa_{u,d}$ are the vev's of the up and down standard model
type Higgs fields in the $\phi(2,2,0)$ multiplet and $v_{u,d}$ are the
corresponding vevs for the same doublets in $\Sigma(2,2,15)$.
Note that there are 13 parameters in the above equations and there are 13
inputs (six quark masses, three lepton masses and three quark mixing
angles and weak scale). Thus all parameters of the model that go into
fermion masses are determined.

 To determine the light neutrino masses, we use the seesaw
formula in Eq. (5), i.e.
\begin{eqnarray}
{ M}_\nu~=~f_\nu v_L-h^T_\nu f^{-1}_{\nu}h_\nu
\left(\frac{v^2_{wk}}{v_R}\right).\label{type2}
\end{eqnarray}
Now we note from Eq.(8), that there is a sum rule relating the
charged lepton and quark mass matrices i.e.
\begin{eqnarray}
k \tilde{M}_\ell = \tilde{M}_u + r \tilde{M}_d
\end{eqnarray}
where $m_3\tilde{M}~=~M$ ($m_3$ being the mass of the third generation
fermion. Fitting the $\tau$ and $\mu$ masses implies that $k,r$ are of
order one. Therefore, the form of the charged lepton mass matrix is
roughly of the same form as the quark mass matrices. Therefore the
argument of the previous section (i.e. exact $SU(4)_c$ case applies here
too and leads to a ``large'' $\theta_{13}$. 

This model grand unifies to a class of minimal R-parity
conserving SO(10) model\cite{minimal} discussed
recently\cite{babu,others1,goran,goh,mimura,fukuyama}. The
$\phi(2,2,1)$ and 
$\Sigma(2,2,15)$ multiplets are embedded into the {\bf 10} and {\bf 126}
dimensional representations of SO(10). An interesting advantage of the
SO(10) unification noted in \cite{babu} is that the triplets
$\Delta\oplus\Delta^c$ along with $\Sigma$ become part of the {\bf 126}
Higgs representation. As a result, we get $f~=~f_\nu$ making the model
quite predictive in the neutrino sector. 
The model leaves R-parity as an automatic symmetry of the low energy
Lagrangian leading to a  naturally stable dark matter in this case.

The model also leads naturally to large neutrino mixing angles without the
need for any fine tuning of right handed neutrino masses as in the model
discussed in the previous section. To see this happens, note that as
already noted earlier, any theory
with asymptotic parity symmetry
leads to type II seesaw formula. 
When the triplet term in the type II seesaw dominates the neutrino mass,
 we have the neutrino mass matrix ${ M}_\nu \propto f$,
where $f$ matrix is the {\bf 126} coupling to fermions discussed earlier.
Using the above equations, one can derive the following
sumrule (sumrule was already noted in the third reference of Ref.[7].).
\begin{eqnarray}
{ M}_\nu~=~ c (M_d - M_\ell)
\label{key}
\end{eqnarray}
where numerically $c\approx 10^{-9}$ GeV.
To see how this leads to large atmospheric and solar mixing,
let us work in the basis where the down quark mass matrix is diagonal. All
the quark mixing effects are then in the up quark mass matrix i.e.
$M_u~=~U^T_{CKM}M^d_u U_{CKM}$. As already noted the minimality of the
Higgs content leads to the following sumrule among the mass matrices:
\begin{eqnarray}
k \tilde{M}_{\ell}~=~r\tilde{ M}_d +\tilde{ M}_u
\end{eqnarray}
where the tilde denotes the fact that we have made the mass matrices
dimensionless
by dividing them by the heaviest mass of the species i.e. up quark mass
matrix by $m_t$, down quark mass matrix by $m_b$ etc. $k,r$ are functions
of the symmetry breaking parameters of the model.
Using the Wolfenstein parameterization for quark mixings, we can conclude
that that we have
\begin{eqnarray}
M_{d,\ell}~\approx ~m_{b,\tau}\left(\begin{array}{ccc}\lambda^4 &
\lambda^3
&\lambda^3\\ \lambda^3 & \lambda^2& \lambda^2 \\ \lambda^3 & \lambda^2 &
1\end{array}\right)
\end{eqnarray}
where $\lambda \sim 0.22$ and the matrix elements are supposed to give
only the approximate order of magnitude.

An important consequence of the relation between the
charged lepton and the quark mass matrices in Eq. (10) is that the charged
lepton
contribution to the neutrino mixing matrix i.e. $U_\ell \simeq {\bf 1} +
O(\lambda)$ or close to identity matrix. As a result
the neutrino mixing matrix
is given by $U_{PMNS}~=~U^{\dagger}_\ell U_\nu \simeq U_\nu$, since in
$U_\ell$, all mixing angles are small. This satisfies one the criteria 
listed in the introduction. Thus the dominant
contribution to large mixings will come from $U_\nu$, which in turn
will be dictated by the sum rule in Eq. (11). Let us now see how how this
comes about.

As we extrapolate the quark
masses to the GUT scale, due to the fact that $m_b-m_\tau \approx
m_{\tau}\lambda^2$ for a range of values of tan$\beta$, the
neutrino mass matrix
$M_\nu~=c(M_d-M_\ell)$ takes roughly the form
\begin{eqnarray}
M_{\nu}~=c(M_d-M_\ell)\approx ~m_0\left(\begin{array}{ccc}\lambda^3 &
\lambda^3
&\lambda^3\\ \lambda^3 & \lambda^2 & \lambda^2 \\ \lambda^3 & \lambda^2
& \lambda^2\end{array}\right)
\end{eqnarray}
This mass matrix is in the form discussed in Eq. (3) (when $\lambda$ is 
factored out in Eq. (3)). 
It is then easy to see that both the $\theta_{12}$
(solar angle) and $\theta_{23}$ (the atmospheric angle) are now large. The
detailed magnitudes of these angles of course depend on the details of the
quark masses at the GUT scale. Using the extrapolated values of the quark
masses and mixing angles to the GUT scale, the
predictions of this model for various oscillation parameters are given in
Ref.[9]. The predictions for the
solar and atmospheric mixing angles fall within 3 $\sigma$ range of the
present central values. Specifically the
prediction for $U_{e3}$ (see Fig. 1) can be tested in various reactor 
experiments\cite{reactor} and at MINOS as well as
other planned Long Base Line neutrino experiments such as Numi-Off-Axis 
(NoVA), JPARC etc.\cite{bnl}.

\begin{figure}
\begin{center}
\epsfxsize8cm\epsffile{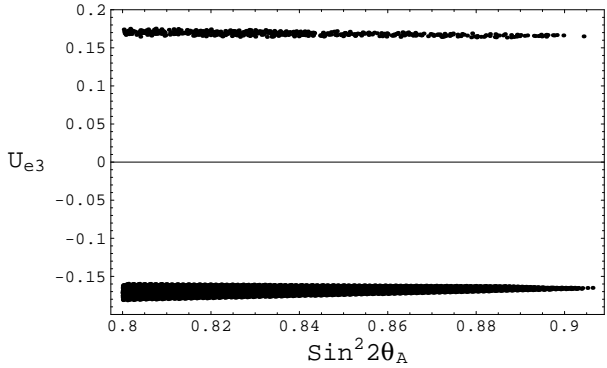}
\caption{
The figure shows the predictions of the minimal SO(10) model for
$sin^22\theta_{A}$
and $U_{e3}$ for the allowed range of parameters in the model. Note that
$U_{e3}$ is very close to the upper limit allowed by the existing reactor
experiments.
\label{fig:cstr3}}
\end{center}
\end{figure}

There is a simple explanation of why the $U_{e3}$ comes out to be
large. This can also be seen from the mass sumrule in
Eq.\ref{key}. Roughly, for a matrix with hierarchical eigen values as is
the case here, the mixing angle $tan 2\theta_{13}\sim
\frac{M_{\nu,13}}{M_{\nu,33}}\simeq \frac{\lambda^3
m_\tau}{m_b(M_U)-m_\tau(M_U)}$. Since to get large mixings, we need
$m_b(M_U)-m_\tau(M_U)\simeq m_\tau \lambda^2$, we see that $U_{e3}\simeq
\lambda$ upto a factor of order one. Indeed the detailed calculations lead
to $0.16$ which is not far from this value.

The model as discussed above does not have CP violation. One way to 
accomodate CP violation is to include {\bf 120} dimensional Higgs 
multiplet into the theory\cite{mimura,bertolini}. By an appropriate choice 
of CP symmetry, one can choose the {\bf 10} and {\bf 126} couplings to be 
real whereas the {\bf 120} coupling is imaginary. This not only introduces 
CP phases into the theory so that one gets CKM CP violation at the weak 
scale; it also provides a solution to the SUSY CP problem as well as 
possibly to the strong CP problem. Despite the presence of the extra Higgs 
multiplet, which brings in three new parameters, the model is still 
predictive for $\theta_{13}$ and in fact one gets a lower limit for 
$\theta_{13} \geq 0.08-0.1$ (Fig 2 below). 

\begin{figure}[h]
\begin{center}
\includegraphics{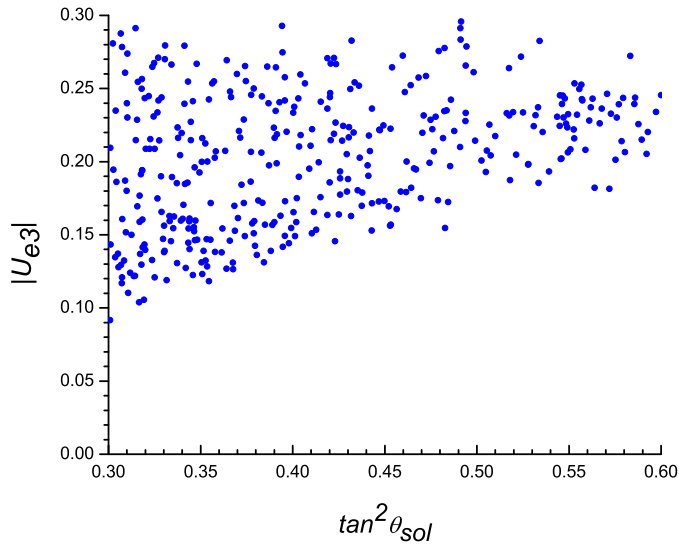}
\caption{$\theta_{13}$ for the minimal SO(10) model with CKM CP 
violation. Scatter of points corresponds to different allowed quark mass
values. Note that the smallest value is around $0.08$. }
\end{center}
\end{figure}

\subsection{Radiative generation of large mixings with quark-lepton 
unification}
As alluded before, type II seesaw liberates the neutrinos from obeying
normal generational hierarchy and instead could easily givbe a 
quasi-degenerate mass spectrum. This provides a new mechanism for 
understanding the large
mixings. The basic idea is that since  at the seesaw
scale, one expects quark-lepton unification to be a good symmetry, we 
expect all mixings angles (i.e. both quark as well as lepton) to be small.
 Since the observed neutrino mixings are the weak scale
observables, one must extrapolate\cite{babu1} the seesaw scale mass
matrices to the weak scale and recalculate the mixing angles.
The extrapolation formula is ${ M}_{\nu}(M_Z)~=~ {\bf I}{ M_{\nu}}
(v_R) {\bf I}$
where ${\bf I}_{\alpha \alpha}~=~
\left(1-\frac{h^2_{\alpha}}{16\pi^2}\right)$.
Note that since $h_{\alpha}= \sqrt{2}m_{\alpha}/v_{wk}$ ($\alpha$ being
the charged lepton index), in the extrapolation only the $\tau$-lepton
makes a difference. In the MSSM, this increases the ${ M}_{\tau\tau}$
entry of the neutrino mass matrix and essentially leaves the others
unchanged. It was shown\cite{balaji} that if the muon and the tau
neutrinos are nearly degenerate but not degenerate enough in mass at the 
seesaw scale and have same CP eigenvalue, then
the radiative corrections can become large enough so that at the weak
scale the two diagonal elements of ${ M}_{\nu}$
 become much more degenerate. This leads to an enhancement of the
mixing angle to its almost maximal value.
This can be seen from the renormalization group equations when they
are written in the mass basis\cite{casas}. Denoting the mixing angles as
$\theta_{ij}$ where $i,j$ stand for generations, the equations are:
\noindent
\begin{eqnarray}
\dsodt&=&-F_{\tau}{c_{23}}^2\left(
-s_{12}U_{\tau1}D_{31}+c_{12}U_{\tau2}D_{32}
\right),\label{eq3}\\
\dstdt&=&-F_{\tau}c_{23}{c_{13}}^2\left(
c_{12}U_{\tau1}D_{31}+s_{12}U_{\tau2}D_{32}
\right),\label{eq4}\\
\dsthdt&=&-F_{\tau}c_{12}\left(c_{23}s_{13}s_{12}U_{\tau1}
D_{31}-c_{23}s_{13}c_{12}U_{\tau2}D_{32}\right.\nonumber \\
&&\left.+U_{\tau1}U_{\tau2}D_{21}\right).\label{eq5}\end{eqnarray}
\noindent
where $D_{ij}={\left(m_i+m_j)\right)/\left(m_i-m_j\right)}$ and
$U_{\tau 1,2,3}$ are functions of the neutrino mixings angles. The
presence
of $(m_i-m_j)$ in the denominator makes it clear that as $m_i\simeq m_j$,
that particular coefficient becomes large and as we extrapolate from the
GUT scale to the weak scale, small mixing angles at GUT scale become large
at the weak scale.

It has been shown recently that indeed such a mechanism for understanding
large mixings can
work for three generations\cite{par}.  The basic idea of Ref.\cite{par} is
to identify the 
neutrino mixing angles with the corresponding quark mixings at the
seesaw scaleand assume quasi-degenerate neutrinos. This can be obtailed in 
models with type II seesaw mechanism and $SU(4)_c$ gauge symmetry.
Then by the mechanism of rediative magnification discussed above, the weak
scale solar
and atmospheric angles get magnified to the desired level while due to the
extreme smallness of $V_{ub}$, the magnified value of $U_{e3}$ remains
within its present upper limit. 

As noted, such a situation can naturally arise in a
parity symmetric model with quark-lepton unification  group $G_{224}$
provided one uses both the
terms in the type II seesaw formula (Eq. \ref{type2}) and use a symmetry
that leades to $f_\nu~=f_0 {\bf I}$ where ${\bf I}$ is the identity
matrix. The first term (triplet vev term) then provides the common mass 
for neutrinos and the standard seesaw term provides the mass splittings.
Using the Higgs fields i.e. $\phi(2,2,1)$ and
$\Sigma(2,2,15)$, one can get a realistic quark spectrum while keeping the
neutrino and quark mixing angles identical at the seesaw scale. They are
then  extrapolated down to the weak scale using the supersymmetric
renormnalization group extrapolation\cite{babu1}.
 In figure 3, we show
the evolution of the mixing angles to the weak scale.
 \begin{figure}
\epsfxsize=8.5cm
\epsfbox{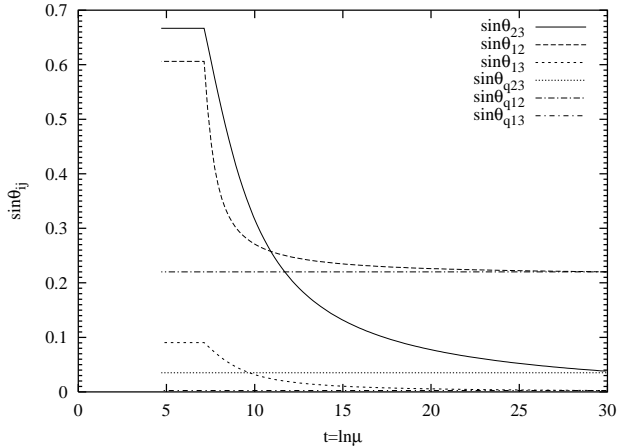}
\caption{Radiative magnification of small quark-like neutrino mixings at
the see-saw scale to bilarge values at low energies. The solid, dashed and
dotted lines represent
$\sin\theta_{23}$, $\sin\theta_{13}$, and $\sin\theta_{12}$,
respectively.}
\label{fig1}
\end{figure}
A requirement for this scenario to work is that the common mass of
neutrinos must be larger than $0.1$ eV, a result that can be tested in
neutrinoless double beta experiments.

\subsection{Quark-lepton complementarity and large solar mixing}
There has been a recent suggestion\cite{raidal} that perhaps the large
but not maximal solar mixing angle is related to physics of the quark
sector. According to this suggestion, the deviation
from maximality
of the solar mixing may be related to the quark mixing angle
$\theta_C\equiv \theta^{q}_{12}$ and is based on the
observation that the mixing angle responsible
for solar neutrino oscillations, $\theta_{\odot}\equiv \theta^\nu_{12}$
satisfies an interesting
complementarity relation with the corresponding angle in the quark sector
$\theta_{Cabibbo}\equiv \theta^q_{12}$ i.e. 
\begin{eqnarray}
\theta^\nu_{12}+\theta^q_{12}
\simeq \pi/4,
\end{eqnarray}
 which seems to be quite well satisfied by present data.
While it
is quite possible that this relation is purely accidental or due to some
other dynamical effects, it is interesting
to pursue the possibility that there is a deep meaning behind it
and see where it leads. It has been shown in a recent paper that if
Nature is quark lepton unified at high scale, then a relation between
$\theta^\nu_{12}$ and $\theta^q_{12}$ can be obtained in a natural manner
provided the neutrinos obey the inverse hierarchy\cite{fram}. This model 
gives a variation of the quark-lepton complementarity relation (Eq. 
(18)) and predicts
$sin^2\theta_\odot\simeq 0.34$ which agrees with present data at the
2$\sigma$ level. It also predicts a large $\theta_{13}\sim 0.18$, both of
which are predictions that can be tested experimentally in the near
future. There are other corrections which affect the value of the solar 
mixing angle\cite{others} e.g. threshold corrections that can bring the 
value closer to the present central value. We also note that an operator 
analysis of QLC has recently been performed in the context of grand 
unification models with quark-lepton unification\cite{king}, where it has 
been pointed out that obtaining QLC necessarily imples $\theta_{13}\simeq 
\theta_C$. This prediction is also in line with the general hypothesis of 
this paper.

\section{$\mu-\tau$ interchange symmetry and small $\theta_{13}$}
A question posed by the above discussions is the following: suppose 
$\theta_{13}$ is found to be below the ``benchmark'' value of $0.04$
or so; what does this imply ? It is clear that obtaining 
such small values in the context of quark lepton unified theories would 
require fine adjustment among parameters. Typically in physics, when a 
small parameter requires fine tuning, that is an indication of an 
underlying symmetry. In the case of $\theta_{13}$, the symmetry turns out 
to be a simple interchange symmetry between $\mu$ and $\tau$ generations 
in the neutrino sector.
The motivation to suspect the existence of such a symmetry is the near
maximal value for the atmospheric mixing angle. To see this note that in 
the neutrino
mass matrix given in Eq. (3), if we set $a=1$ and $b=d$, then the mass
matrix is invariant under the interchange of $\mu-\tau$ labels.
It is then easy to see by
diagonalizing this mass matrix that it leads to $\theta_A=\pi/4$ and
$\theta_{13}=0$. Thus we conclude that a very small value of $\theta_{13}$
can be understood if the $\mu-\tau$ symmetry is very nearly exact much the
same way one hopes to understand a small value of electron mass as a
consequence of chiral symmetry breaking since in the limit of exact chiral
symmetry, $m_e=0$.

In fact if one breaks $\mu-\tau$ symmetry in the $\mu-\tau$ sector by
making the neutrino mass matrix to take the following form:
\begin{eqnarray}
{ M}_\nu~=~\frac{\sqrt{\Delta
m^2_A}}{2}\left(\begin{array}{ccc}c\epsilon
&b\epsilon &b\epsilon\\ b\epsilon & 1+\epsilon & -1 \\
d\epsilon & -1 & 1+\epsilon\end{array}\right)
\end{eqnarray}
 then it is easy to see that, one gets a nonzero $\theta_{13}$ given by:
\begin{eqnarray}
\theta_{13}\simeq
\frac{1}{4\sqrt{2}}\epsilon^2d(1-a)
\end{eqnarray} where
$\epsilon \simeq
\frac{4}{[c+(1+a)/2]+\sqrt{[c-(1+a)/2]^2+8d^2}}\sqrt{\frac{\Delta
m^2_\odot}{\Delta m^2_A}}$.
Thus the existence of $\mu-\tau$ symmetry allows us to theoretically
understand a truly small $\theta_{13}$.
An important characteristic of this model is that
there is a strong correlation between the value of $\theta_{13}$ and 
deviation of the atmospheric mixing angle from its maximal value. 
In the 
last figure of this article\cite{yu}, we display the correlation as a 
scatter plot.

\begin{figure}[h]
\begin{center}
\epsfxsize8cm\epsffile{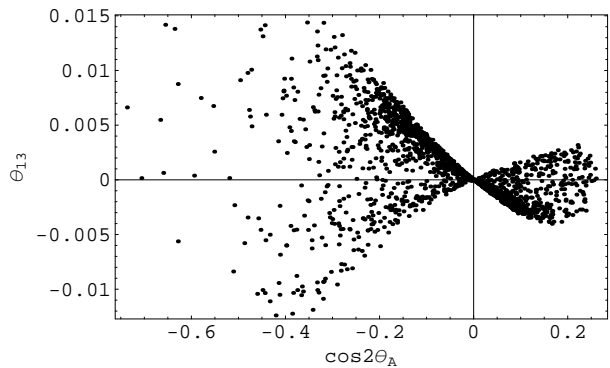}
\caption{Departure from $\mu-\tau$ symmetry and correlation
between $\theta_{13}$ and $\theta_A$. }
\end{center}
\end{figure}

In conclusion, I have argued that if the upper limit on the neutrino
mixing parameter $\theta_{13}$ goes below around $0.04$ or so,
 then this would strongly suggest that there is no quark
lepton unification such as those based on the groups $SU(2)_L\times
SU(2)_R\times SU(4)_c$ or SO(10) at high scale. This would be a
significant step in our search for the ultimate unified theory of forces
and matter and therefore provides very strong motivation for high
precision experimental search for $\theta_{13}$. Two exceptions to this 
conclusion are that: (i) there exist sterile neutrinos and/or (ii) the 
lepton mixings receive dominant contributions from the charged lepton 
sector\cite{albright}.

 This work is supported by the National Science Foundation
Grant No. PHY-0354401 and also partly by GRDF No.3305. I would like to 
thank Stefan Antusch, B. Dutta, P. Frampton, Steve King, Y. Mimura, Salah 
Nasri, G. Senjanovic and Hai-bo Yu for discussions.

\end{document}